\documentclass[english,twocolumn,showpacs,preprintnumbers,amsmath,amssymb,pra]{revtex4}
\usepackage[T1]{fontenc}
\usepackage[latin1]{inputenc}
\usepackage{prettyref}
\usepackage{amsmath}
\usepackage{graphicx}
\usepackage{amssymb}

\makeatletter

\newcommand{\lyxdot}{.}

\makeatletter

\usepackage{bm}

\makeatletter

\renewcommand{\Re}[1]{\mathrm{Re\,}}
\renewcommand{\Im}[1]{\mathrm{Im\,}}
\renewcommand{\vec}[1]{\bm{#1}}

\makeatother

\usepackage{babel}
\makeatother
\begin{document}

\title{Non-BCS superfluidity in trapped ultracold Fermi gases}

\author{L. M. Jensen$^{\text{1}}$, J. Kinnunen$^{\text{1,2}}$ and P. T\"{o}rm\"{a}$^{\text{1}}$}

\affiliation{$^{1}$Department of Physics, Nanoscience Center, P.O.Box 35, FIN-40014
University of Jyv\"{a}syl\"{a}, Finland\\
$^{2}$JILA and Department of Physics, University of Colorado at Boulder,
CO 80309-0440, USA }


\begin{abstract}
Superconductivity and superfluidity of fermions require, within the
BCS theory, matching of the Fermi energies of the two interacting
Fermion species. Difference in the number densities of the two species
leads either to a normal state, to phase separation, or - potentially
- to exotic forms of superfluidity such as FFLO-state, Sarma state
or breached pair state. We consider ultracold Fermi gases with polarization,
i.e.\ spin-density imbalance. We show that, due to the gases being
trapped and isolated from the environment in terms of particle exchange,
exotic forms of superfluidity appear as a shell around the BCS-superfluid
core of the gas and, for large density imbalance, in the  core as
well. We obtain these results by describing the effect of the trapping
potential by using the Bogoliubov-de Gennes equations. For comparison
to experiments, we calculate also the condensate fraction, and show
that,  in the center of the trap, a polarized superfluid leads to
a small dip in the central density difference. We compare the results
to those given by local density approximation and find qualitatively
different behavior. 
\end{abstract}
\maketitle

\affiliation{$^{\text{1}}$Department of Physics, Nanoscience Center, P.O.Box
35, FIN-40014 University of Jyväskylä, Finland \\
$^{\text{2}}$JILA and Department of Physics, University of Colorado
at Boulder, CO 80309-0440, USA}

\section{Introduction}

There are several suggestions of non-BCS superfluidity for fermion
systems with spin-population imbalance, i.e. with the polarization
$P=(N_{\uparrow}-N_{\downarrow})/(N_{\uparrow}+N_{\downarrow})\neq0$
where $N_{\sigma}$ are the particle numbers of the (pseudo)spins.
The FFLO-state \cite{Fulde1964,Larkin1965}, the Sarma state \cite{Sarma1963}
and the breached pair (BP) state \cite{Liu2003} all appear as extremal
points of the mean-field energy of the system. Such superfluidity
is of interest e.g. in condensed matter, high-energy and nuclear physics
\cite{Casalbuoni2004}, but firm experimental evidence is lacking.
With the recently realized superfluids of alkali Fermi gases, the
first studies of density-imbalanced gases \cite{Zwierlein2005c,Partridge2005,Zwierlein2006a,Partridge2006c}
have shown that these systems offer unprecedented opportunities for
investigating this question. The FFLO-state is predicted to appear
in a narrow parameter window in several systems \cite{Sedrakian2005,Mizushima2005a,Castorina2005,Sheehy2006},
and the Sarma/BP state has been shown to be unstable under many conditions
\cite{Gubankova2003,Forbes2005}. The existence of these exotic forms
of superfluidity is thus an intriguingly subtle question and requires
careful analysis, taking into account the specific features of the
physical system. In this article, we demonstrate the important consequences
of such features in case of trapped ultracold alkali gases. First,
in ultracold gases, the physical system under study is isolated from
the environment in terms of particle exchange. This could be contrasted
to an electronic system where external voltage fixes the chemical
potential by allowing particle exchange between the system of interest
and the environment. Second, the gas is trapped by an inhomogeneous
potential (often of harmonic form). The second issue leads to phase
separation of superfluid and normal phases, as shown by a series of
experiments \cite{Zwierlein2005c,Partridge2005,Zwierlein2006a,Zwierlein2006c,Shin2006a,Partridge2006b}
and theoretical studies \cite{Bedaque2003,Sheehy2006,Pieri2006a,Kinnunen2006a,Yi2006a,Haque2006,Chevy2006,deSilva2006a,Chien2006b,Gubbels2006b}.
The finiteness and harmonic confinement of the system leads, as we
show in this article, to stabilization of exotic forms of superfluidity
in a shell surrounding a BCS-superfluid core, and eventually in the
core itself. 

Ultracold Fermi gases offer the possibility of realizing the BCS-BEC
crossover with the use of Feshbach resonances. At the resonance point,
the system is a strongly interacting Fermi superfluid, and on different
sides away from the resonance it is either a BEC of molecules or a
weakly interacting Fermi gas realizing a BCS superfluid. Condensation
of molecules, fermion pairs, pairing gap and the crossover behavior
have been studied by a series of experiments and vortices confirming
superfluidity have been created, for a review see for instance~\cite{Grimm2005}.
Recently, seminal experiments studying density imbalanced (nonzero
polarization $P$) Fermi gases were done~\cite{Zwierlein2005c,Partridge2005,Zwierlein2006a}.
The analysis presented in \cite{Zwierlein2005c,Shin2006a,Zwierlein2006c}
combined the study of vortex patterns, density profiles (in \cite{Shin2006a}
3D reconstructed) and condensate fractions, and thereby showed that,
 in the trapped system, a superfluid core (with equal densities of
the components $\uparrow$ and $\downarrow$) appears in the  center
of the trap and the excess atoms of the majority component ($\uparrow$)
tend to be located on the edges of the trap.  In the following, the
BCS paired superfluid (SF) implies equal local densities, whereas
the polarized superfluid (PS) at zero temperature denotes exotic forms
of superfluidity with non-zero local density difference of the $\uparrow$
and $\downarrow$ components. In addition, the normal state (N) shell
surrounding the condensate can either be partially polarized or fully
polarized depending on whether the minor component is present or not. 

In connection to the experiments \cite{Zwierlein2005c,Partridge2005,Zwierlein2006a},
several authors have analyzed the trapped, polarized Fermi gas using
the local density approximation (within mean-field theory) \cite{Sheehy2006,Pieri2006a,deSilva2006a,Chevy2006,Yi2006a,Haque2006,Chien2006b,Gubbels2006b,beyond}.
In these works, the problem was solved at each spatial point, applying
a local chemical potential, and the stability of the solutions at
that point was determined by their grand potential energies. This
leads, at zero temperature, to the exclusion of Sarma/BP state which
is known to be a maximum point of the grand potential for fixed chemical
potentials. This gives the following qualitative picture: a superfluid
(SF) core, with equal densities, appears at the center of the trap,
and is surrounded by a normal state (N) shell (only on the BEC side
of the Feshbach resonance a coexistence of molecular BEC and free
atoms was seen, as obviously expected since molecule formation does
not require matching Fermi energies). This is, however, a rather approximative
description of the system because LDA assumes a smooth variation of
the density. Since the studies \cite{Sheehy2006,deSilva2006a,Chevy2006,Yi2006a,Haque2006,Chien2006b}
imply a step in the density at the interface of the SF core and the
N shell one can question the validity of LDA at that boundary, as
also remarked by many of the authors of \cite{Sheehy2006,deSilva2006a,Chevy2006,Yi2006a,Haque2006,Chien2006b}
and recent beyond-LDA studies \cite{Partridge2006b,Imambekov2006,deSilva2006b}.
\begin{figure}
\begin{centering}\includegraphics[width=7cm]{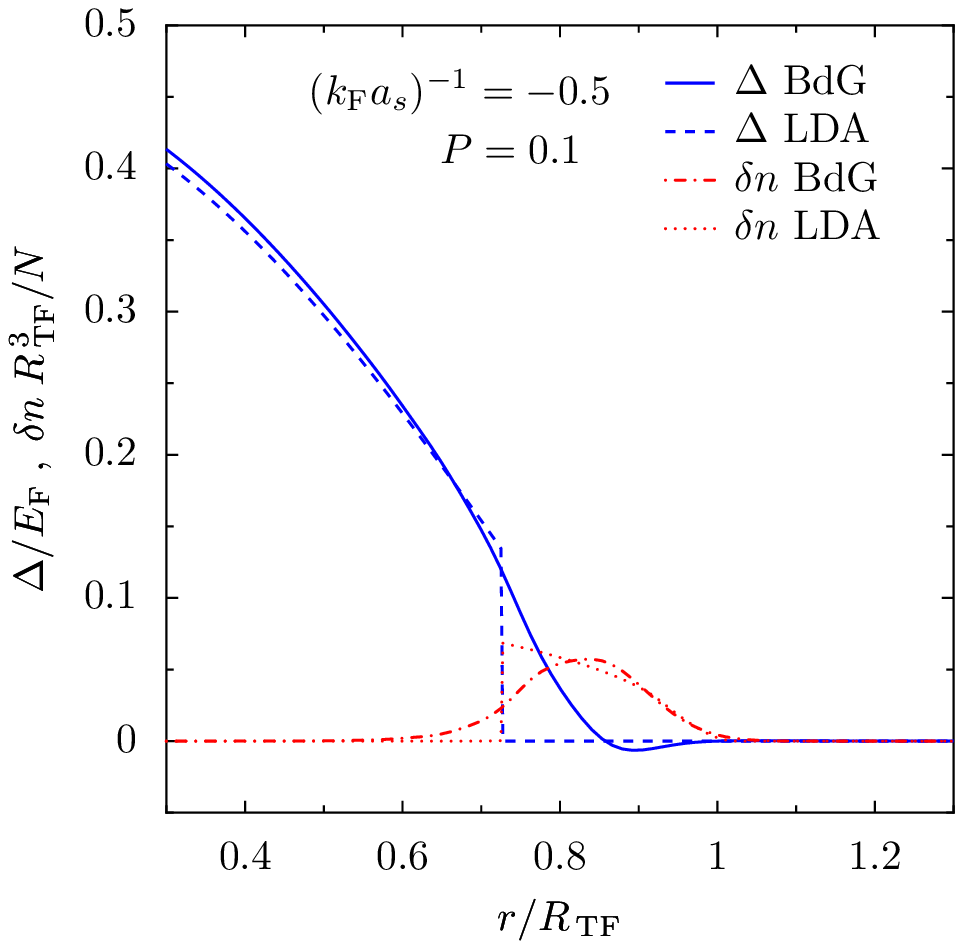}\par\end{centering}

\caption{(color online) The gap $\Delta(r)$ and density difference $\delta n=n_{\uparrow}(r)$
-- $n_{\downarrow}(r)$ profiles for the LDA and BdG calculations.
\label{fig:comp-zero-T}}
\end{figure}

The existence of a PS for trapped, strongly interacting gases was
indicated already in our earlier study \cite{Kinnunen2006a}, where
the treatment of the trapping potential by Bogoliubov-de Gennes (BdG)
equations revealed oscillations of the order parameter in an area
located between the SF core and the N shell. Such oscillations resemble
the nonuniform order parameter associated with the FFLO state, therefore
we refer to these oscillations as \char`\"{}FFLO-type state\char`\"{}.
In this article, we show that the BdG analysis predicts such polarized
superfluid in trapped Fermi gases not only as a shell effect but,
for large polarizations, as a feature that extends through the whole
system. We confirm that superfluidity and finite local density difference
indeed co-exist in the center of the trap by calculating the condensate
fraction, central gap and the core polarization. In previous works
using BdG for trapped gases \cite{Castorina2005,Mizushima2005a,Kinnunen2006a,Machida2006}
such an analysis, confirming the existence of a polarized superfluid
in the center, was not performed. Moreover, the works \cite{Castorina2005,Mizushima2005a,Machida2006}
considered the weak interaction limit whereas we have extended the
BdG calculation to the unitarity regime and actually consider the
whole crossover from the BCS to BEC side. We analyze the dependence
of the oscillations on the system size (particle number) and discuss
the connection to phenomena occurring in superconductor - ferromagnet
interfaces. For comparison, we also perform calculations using LDA.
The LDA analysis predicts a polarized superfluid only at finite temperature,
not at zero temperature like the BdG calculation. Therefore our results
show that, in trapped Fermi gases, LDA has to be applied with care;
beyond LDA approaches are not only needed for describing the interfaces
of the system properly but, in some cases, features not predicted
by LDA may become dominant.

\begin{figure}
\begin{centering}\includegraphics[width=7cm]{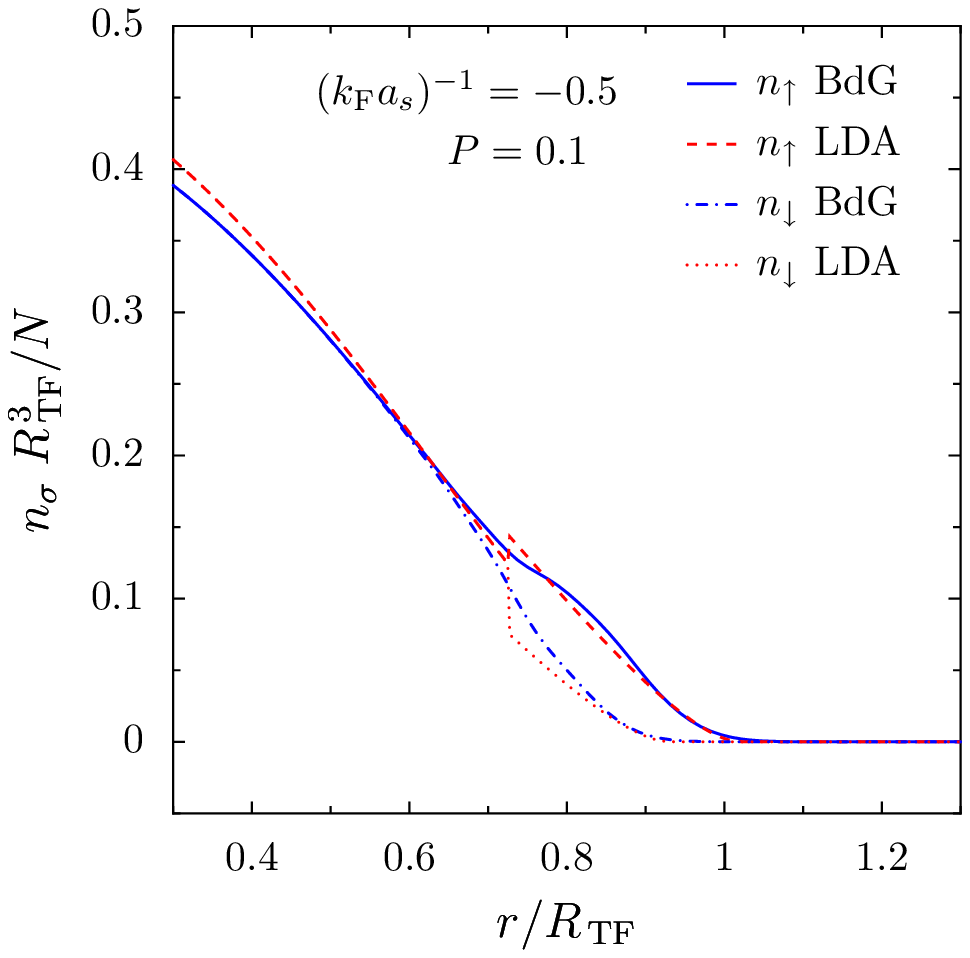}\par\end{centering}

\caption{(color online) The density profiles of the $n_{\uparrow}(r)$ for
BdG (solid) and LDA (dashed), and $n_{\downarrow}(r)$ for BdG (dot-dashed)
and LDA (dotted). \label{fig:dens-comp-zero-T}}
\end{figure}
This article is structured as follows. In section II. we present a
review of the Bogoliubov - de Gennes (BdG) approach for describing
pairing at the mean field level. We discuss the use of local density
approximation, and the expansion in harmonic oscillator states as
special cases of this general scheme. In the rest of the paper, we
use the following terminology: BdG refers to the case where harmonic
oscillator state basis has been used, and LDA to the use of local
density approximation. The results are presented in section II. The
appearance of the FFLO-type oscillations is discussed in subsection
II.A and the case of large polarizations, when such oscillations span
the whole system, is discussed in section II.B. We also analyze the
dependence of these features on the interaction strength through the
BCS-BEC crossover (section II.A.1.) and on the system size (atom number)
(section II.A.2). The condensate fraction is calculated in section
II.C. and the contributions of different harmonic oscillator states
to the pairing are discussed in section II.D. Conclusions and discussion
are presented in section III. %
\begin{figure}
\begin{centering}\includegraphics[width=7cm]{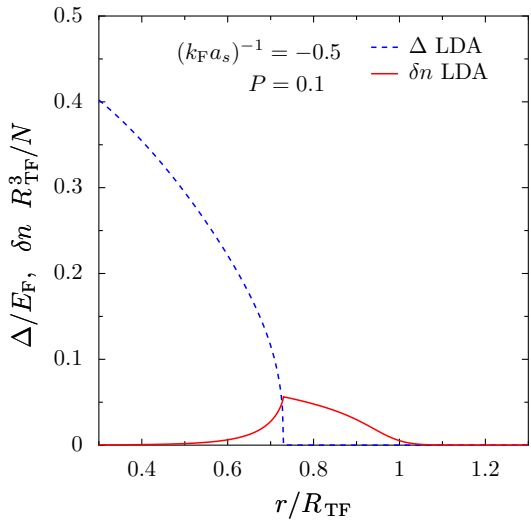}\par\end{centering}

\caption{(color online) The gap and polarization profiles at slightly elevated
temperature $T/T_{\mathrm{F}}=0.05$ for the LDA calculations and
at low polarization. In contrast to the zero temperature case the
thermal excitation in the BCS core starts to show up as a tail of
the density difference into the core. \label{fig:comp-finite-T}}
\end{figure}

\section{Bogoliubov - de Gennes approach}

In order to properly account for the inhomogeneity due to the presence
of the trap potential we will in the following use the Bogoliubov-de
Gennes equations \cite{deGennes1999}, which in a more general context
are also known as the selfconsistent Hartree-Fock-Bogoliubov equations.
The imbalanced two-component Fermi gas is described by the grand canonical
Hamiltonian \begin{eqnarray}
H & = & \sum_{\sigma}\int d^{3}\vec{r}\ \hat{\Psi}_{\sigma}^{\dagger}(\vec{r})\left(-\frac{\hbar^{2}\nabla^{2}}{2m}-\mu_{\sigma}+V(\vec{r})\right)\hat{\Psi}_{\sigma}(\vec{r})\nonumber \\
 & + & \int d^{3}\vec{r}d^{3}\vec{r}^{\prime}\ \hat{\Psi}_{\uparrow}^{\dagger}(\vec{r})\hat{\Psi}_{\downarrow}^{\dagger}(\vec{r})U(\vec{r}-\vec{r}^{\prime})\hat{\Psi}_{\downarrow}(\vec{r})\hat{\Psi}_{\uparrow}(\vec{r}),\end{eqnarray}
where $\hat{\Psi}_{\sigma}(\vec{r}),\hat{\Psi}_{\sigma}^{\dagger}(\vec{r})$
are the real space annihilation and creation operators for an atom
with spin $\sigma$ at position $\vec{r}$, $\mu_{\sigma}$ is the
chemical potential for the components $\sigma,$ $V(\vec{r})=\frac{1}{2}m\omega^{2}r^{2}$
is the external (isotropic and harmonic) trapping potential, and $U(\vec{r})=U\delta(\vec{r})$
is the interatomic atom-atom contact interaction potential. We apply
the contact potential interaction and the mean-field approximation
\begin{eqnarray}
U\hat{\Psi}_{\uparrow}^{\dagger}(\vec{r})\hat{\Psi}_{\downarrow}^{\dagger}(\vec{r})\hat{\Psi}_{\downarrow}(\vec{r})\hat{\Psi}_{\uparrow}(\vec{r}) & \approx & -\frac{\left|\Delta(r)\right|^{2}}{U}-Un_{\uparrow}(\vec{r})n_{\downarrow}(\vec{r})\nonumber \\
 &  & \hspace{-3.5cm}+\, Un_{\uparrow}(\vec{r})\hat{\Psi}_{\downarrow}^{\dagger}(\vec{r})\hat{\Psi}_{\downarrow}(\vec{r})+Un_{\downarrow}(\vec{r})\hat{\Psi}_{\uparrow}^{\dagger}(\vec{r})\hat{\Psi}_{\uparrow}(\vec{r})\nonumber \\
 &  & \hspace{-3.5cm}+\,\Delta(\vec{r})\hat{\Psi}_{\uparrow}^{\dagger}(\vec{r})\hat{\Psi}_{\downarrow}^{\dagger}(\vec{r})+\Delta^{\ast}(\vec{r})\hat{\Psi}_{\downarrow}(\vec{r})\hat{\Psi}_{\uparrow}(\vec{r}),\end{eqnarray}
where $\Delta(\vec{r})=\langle\hat{\Psi}_{\downarrow}(\vec{r})\hat{\Psi}_{\uparrow}(\vec{r})\rangle$
and $n_{\sigma}(\vec{r})=\langle\hat{\Psi}_{\sigma}^{\dagger}(\vec{r})\hat{\Psi}_{\sigma}(\vec{r})\rangle$.
The first two terms are simply numbers, so they are neglected. This
gives the following mean-field Hamiltonian\begin{eqnarray}
H_{\mathrm{MF}} & = & \sum_{\sigma}\int d^{3}\vec{r}\ \hat{\Psi}_{\sigma}^{\dagger}(\vec{r})\left(-\frac{\hbar^{2}\nabla^{2}}{2m}-\mu_{\sigma}+V(\vec{r})\right)\hat{\Psi}_{\sigma}(\vec{r})\nonumber \\
 & + & \sum_{\sigma}\int d^{3}\vec{r}\ Un_{\sigma}(\vec{r})\hat{\Psi}_{\bar{\sigma}}^{\dagger}(\vec{r})\hat{\Psi}_{\bar{\sigma}}(\vec{r})\nonumber \\
 & + & \int d^{3}\vec{r}\left(\Delta(\vec{r})\hat{\Psi}_{\uparrow}^{\dagger}(\vec{r})\hat{\Psi}_{\downarrow}^{\dagger}(\vec{r})+\ \mathrm{H.c.}\right),\label{eq:MFH}\end{eqnarray}
with $\bar{\sigma}$ being the other component of $\sigma.$ The second
line is the mean-field Hartree corrections which represents an asymmetric
shift to the chemical potential $\mu_{\sigma}$ of the gas of $\sigma$
atoms which is proportional to the density of the other component
$n_{\bar{\sigma}}.$ In the Bogoliubov-de Gennes approach one expands
the field operators on an appropriate basis set, dictated by the symmetries
of the non-interacting part of the Hamiltonian and usually characterized
by a set $\{\eta\}$ of good quantum numbers, in order to diagonalize
the Hamiltonian as in the uniform case. In the balanced non-uniform
case and  in the absence of superflow the generalized BCS pairing
is in general between atoms in time-reversed states. The problem can
now be solved by introducing the generalized canonical Bogoliubov-Valatin
transformation to the new fermionic operators $\alpha_{\eta},\alpha_{\eta}^{\dagger}$
which amounts to the expansion \[
\hat{\Psi}_{\sigma}(\vec{r})=\sum_{\eta}u_{\eta}(\vec{r})\hat{\alpha}_{\eta\sigma}-s_{\sigma}v_{\eta}^{\ast}(\vec{r})\hat{\alpha}_{\bar{\eta}\bar{\sigma}}^{\dagger},\]
where the overhead bar designates the quantum numbers of the time-reversed
state for $\eta$ and the other hyperfine spin state for $\sigma,$
and with $s_{\uparrow}=1,s_{\downarrow}=-1.$ We note that in the
analogous case of electronic pairing in superconductors $\bar{\sigma}$
denote the time-reversed spin part of the wavefunction. From the requirement
that the new operators $\alpha_{\eta},\alpha_{\eta}^{\dagger}$ diagonalize
the Hamiltonian \prettyref{eq:MFH} one derives the matrix Bogoliubov-de
Gennes equation \cite{deGennes1999} \begin{equation}
(\hat{H}_{0}\hat{\tau}_{3}-\hat{\Delta}\hat{\tau}_{1})\vec{\varphi}_{\eta}=E_{\eta}\vec{\varphi}_{\eta}(\vec{r}),\label{eq:bdg-eqn}\end{equation}
for the spinor $\vec{\varphi}(\vec{r})\equiv(u_{\eta}(\vec{r}),v_{\eta}(\vec{r}))$,
where $\eta$ denotes a set of appropriate quantum numbers and $u_{\eta},v_{\eta}$
are therefore to be regarded as subspinors, $\hat{H}_{0}$ is the
non-interacting diagonal part of the Hamiltonian, potentially including
a trapping potential and Hartree shifts to the chemical potentials,
$\hat{\Delta}$ is the pairing field part of the Hamiltonian, $E_{\eta}$
is the eigenenergy. The products with the Pauli matrices $\hat{\tau}_{i}$
on the left hand side of Eq. \prettyref{eq:bdg-eqn} are to be understood
as a direct products. The selfconsistent aspect of the method is due
to the fact that the chemical potentials, mean-field Hartree and pairing
fields are to be selfconsistently determined through the gap and number
equations. The details of the BdG calculation for the harmonic trap
eigenstates are presented in Section \prettyref{sub:Harmonic-trap-}.
Next we turn to the discussion of the local density approximation.

\subsection{Local density approximation}

We first consider the local density approximation which is assumed
to be valid for sufficiently large condensates. The starting point
for the LDA calculation is to solve the problem  in the uniform case.
The translational invariance of a uniform superfluid implies that
the plane wave states $\hat{\Psi}(\vec{r)}=\mathcal{V}^{-1/2}\sum_{\vec{k}}e^{i\vec{k}\cdot\vec{r}}a_{\vec{k}}$
can be used to diagonalize the Hamiltonian. The main assumption of
the LDA is that the system is locally homogeneous and therefore we
initially consider the Hamiltonian $H$ in \prettyref{eq:uni-bcs-hamiltonian}
for a uniform system (i.e. $V(r)=0$) with contact interactions, which
in the momentum representation reads \begin{equation}
H=\sum_{\vec{k},\sigma}\varepsilon_{\vec{k}}a_{\vec{k}\sigma}^{\dagger}a_{\vec{k}\sigma}+\frac{U}{\mathcal{V}}\sum_{\vec{k},\vec{k}^{\prime}}a_{\vec{k}\uparrow}^{\dagger}a_{\vec{k}\downarrow}^{\dagger}a_{-\vec{k}^{\prime}\downarrow}a_{\vec{k}^{\prime}\uparrow},\label{eq:uni-bcs-hamiltonian}\end{equation}
 where $a_{\vec{k}\sigma}^{\dagger},a_{\vec{k}\sigma}$ are the creation
and annihilation operators for free atoms with momentum $\hbar\vec{k}$,
and the kinetic energy $\varepsilon_{\vec{k}}=\hbar^{2}k^{2}/(2m)$.
The chemical potentials $\mu_{\sigma}$ are introduced as Lagrange
multipliers for the particle numbers $N_{\sigma}$. We define the
average chemical potential $\mu=(\mu_{\uparrow}+\mu_{\downarrow})/2$
and the imbalance potential $\delta\mu=(\mu_{\uparrow}-\mu_{\downarrow})/2$,
such that $\mu_{\sigma}=\mu+s_{\sigma}\delta\mu,$ with $s_{\sigma}$
defined as above. Within the mean field approximation the Hamiltonian
can be diagonalized by the standard Bogoliubov-Valatin transformation.
In the uniform case the order parameter is the usual $\Delta=\langle a_{\vec{k}\uparrow}a_{-\vec{k}\downarrow}\rangle$
which satisfies the gap equation. Within the semiclassical approximation
the gap $\Delta(r)$ is assumed to depend weakly on the radial position
$r=|\vec{r}|$ and, as is commonly done, we introduce the local chemical
potential $\mu_{\sigma}\to\mu_{\sigma}(r)=\mu_{\sigma}-V(r),$ where
$V(r)$ is the trapping potential. The quasi-particle dispersions
depend on the gap and therefore on the radial position through the
local chemical potentials and the gap profile. In the most general
case the quasi-particle dispersion relation becomes \[
E_{\vec{k}\sigma}(\vec{r})=\frac{1}{2}\left(\xi_{\vec{k}\sigma}(r)-\xi_{\vec{k}\bar{\sigma}}(r)\right)+E_{\vec{k}}(r),\]
where $E_{\vec{k}}(r)=[\xi_{\vec{k}}(r)+\Delta^{2}(r)]^{1/2},$ with
$\xi_{\vec{k}}(r)=\left(\xi_{\vec{k}\sigma}(r)+\xi_{\vec{k}\bar{\sigma}}(r)\right)/2,$
and $\xi_{\vec{k}\sigma}(r)=\hbar^{2}k^{2}/(2m)-\mu_{\sigma}(r).$
As we will assume pairing between atoms with equal mass we get $E_{\vec{k}\sigma}=-s_{\sigma}\delta\mu+E_{\vec{k}},$
and therefore ${2E}_{\vec{k}}=E_{\vec{k}\sigma}+E_{\vec{k}\bar{\sigma}},$
and $\xi_{\vec{k}}=\varepsilon_{\vec{k}}-\mu(r)$ with $\mu,\delta\mu$
defined above. Within LDA, both the gap and the local chemical potentials
depend on the radial position and therefore the system may in general
be locally polarized. We define an averaged Fermi energy scale $E_{\mathrm{F}}$
from the total atom number $N=[E_{\mathrm{F}}/(\hbar\omega)]^{3}/3$,
for a balanced non-interacting Fermi gas in a harmonic trap potential
$V(r)=m\omega^{2}r^{2}/2$, where $\omega$ is the trap frequency.
The characteristic scale of the size of the cloud is the Thomas-Fermi
radius is $R_{\mathrm{TF}}=[2E_{\mathrm{F}}/(m\omega^{2})]^{1/2}$.
Within LDA the local gap equation at position $f$ reads \begin{equation}
1=\frac{U_{\ast}}{\mathcal{V}}\sum_{\vec{k}}\left[\frac{1}{2\varepsilon_{\vec{k}}}-\frac{1-n_{\mathrm{F}}(E_{\vec{k}\uparrow}(r))-n_{\mathrm{F}}(E_{\vec{k}\downarrow}(r))}{2E_{\vec{k}}(r)}\right],\label{eq:gap}\end{equation}
with $U_{\ast}=4\pi\hbar^{2}a_{s}/m$ being the usual regularized
effective interaction strength which replaces the bare interaction
$U,$ and $a_{s}$ is the $s$-wave scattering length, and where we
have regularized the ultraviolet divergence appearing  momentum integrals
arising from unphysical properties of the contact interaction \cite{Fetter1971}.
The gap equation is an implicit equation for the gap profile $\Delta(\vec{r})$
\cite{Perali2003a} which for an isotropic trap is only a function
of the radial position $r=|\vec{r}|$. The number equation and the
density profiles are determined from the thermodynamic relation \begin{eqnarray*}
N_{\sigma} & = & -\frac{\partial\Omega}{\partial\mu_{\sigma}}=\int d^{3}\vec{r}\ n_{\sigma}(r),\end{eqnarray*}
 where the radial distribution for the $\sigma$ component is \[
n_{\sigma}(r)=\frac{1}{\mathcal{V}}\sum_{\vec{k}}\left[u_{\vec{k}}^{2}n_{\mathrm{F}}(E_{\vec{k}\sigma}(r))+v_{\vec{k}}^{2}n_{\mathrm{F}}(-E_{\vec{k}\bar{\sigma}}(r))\right],\]
with $\bar{\sigma}$ denotes the other component of $\sigma.$ The
polarization $P$ is \begin{equation}
P=\frac{N_{\uparrow}-N_{\downarrow}}{N_{\uparrow}+N_{\downarrow}}=\frac{1}{N}\int d^{3}r\ \delta n(r),\end{equation}
 where $\delta n(r)=n_{\uparrow}(r)-n_{\downarrow}(r).$ The equations
for the polarization $P$ and for the minority component $N_{\downarrow}/N=(1-P)/2$
are numerically solved by iteration together with the gap profile
as follows. We first make an initial guess for the values of the chemical
potentials $\mu$ and $\delta\mu$ at fixed coupling $k_{\mathrm{F}}a_{s}.$
The $r$ dependent gap $\Delta(r)$, minor density $n_{\downarrow}(r)$
and polarization density $\delta n(r)$ are then discretized on a
radial grid of sufficient resolution. From the initial values of $\mu$
and $\delta\mu$ and at fixed coupling the gap profile $\Delta(r)$
is calculated by solving the gap equation \prettyref{eq:gap} as a
root problem. The gap profile is then subsequently used to calculate
the densities $n_{\downarrow}(r)$ and $\delta n(r)$ as functions
of $\mu$ and $\delta\mu.$ The minor number and and polarization
equations are then solved as a multidimensional root problem for the
average chemical potential $\mu$ and the bare depairing width $\delta\mu.$
The system of equations are iterated until the gap profile and the
chemical potentials are sufficiently converged and then finally $n_{\uparrow}(r)=\delta n(r)+n_{\downarrow}(r).$
The present method applied here is well suited for considering effects
of the Hartree terms, but we have not included those in the results
presented here.

\subsection{Harmonic trap  eigenstates \label{sub:Harmonic-trap-}}

The Bogoliubov-de Gennes (BdG) approach allows treating the effect
of the harmonic trapping potential exactly. The approach has been
used for polarized Fermi gases \cite{Castorina2005,Mizushima2005a,Kinnunen2006a,Machida2006}
because it avoids some of the problems of a local density approximation
approach. We have used it in our previous work \cite{Kinnunen2006a}
and give in this section a detailed description of the calculations
on which the results in \cite{Kinnunen2006a} and in this article
are based. One particularly important advantage of this approach is
the proper description of interface effects in a phase separated gas,
following from the nonlocal nature of the solution and the wavefunctions.
Eventually, in the limit of a large number of atoms, the LDA and BdG
solutions are expected to become the same. 

The BdG approach used here is a generalization, to the imbalanced
densities case, of the calculation outlined in Ref. \cite{Ohashi2005a}
which was for an unpolarized Fermi gas. For the imbalanced case the
generalization amounts to introducing different chemical potentials
$\mu_{\downarrow}$ and $\mu_{\uparrow}$ for the two species of fermionic
atoms. We now expand the wavefunctions $\hat{\Psi}_{\sigma}(\vec{r})$
 in the basis of the 3D harmonic oscillator eigenstates \begin{equation}
\hat{\Psi}_{\sigma}(\vec{r})=\sum_{nlm}R_{nl}(r)Y_{lm}(\hat{r})\hat{a}_{nlm\sigma},\end{equation}
where the quantum numbers $\{\eta\}\equiv\{ n,l,m\}$ are the radial
quantum number $n$ counting the nodes in the radial function, $l$
is the orbital angular momentum and $m$ is the projected angular
momentum onto the axis of quantization. Here $Y_{lm}(\hat{r})$ are
the spherical harmonics with $\hat{r}=(\theta,\varphi)$ and the radial
part of the wavefunction is \begin{equation}
R_{nl}(r)=\sqrt{2}\left(m\omega\right)^{3/4}\sqrt{\frac{n!}{\left(n+l+1/2\right)!}}e^{-\bar{r}^{2}/2}\bar{r}^{l}L_{n}^{l+1/2}\left(\bar{r}^{l}\right),\end{equation}
where $\bar{r}=r/a_{\mathrm{osc}}$ with the harmonic oscillator length
being $a_{\mathrm{osc}}=[\hbar/(m\omega)]^{1/2},$ and $L_{n}^{l+1/2}(\bar{r}^{2})$
is an associated Laguerre polynomial. The characteristic scale of
the cloud is given by the Thomas-Fermi radius $R_{\mathrm{TF}}^{2}=2E_{\mathrm{F}}/(m\omega^{2})$
and therefore $R_{\mathrm{TF}}=(24N)^{1/6}a_{\mathrm{osc}}$. The
spherical symmetry allows doing the angular integrations, and getting
rid of the $m$-quantum numbers, thereby making the $l$-states $(2l+1)$-fold
degenerate. The expansion yields the following Hamiltonian \begin{eqnarray}
H_{\mathrm{MF}} & = & \sum_{n,l,\sigma}(2l+1)\left(\varepsilon_{nl}-\mu_{\sigma}\right)a_{nl\sigma}^{\dagger}a_{nl\sigma}\nonumber \\
 & + & U\sum_{n,n^{\prime},l,\sigma}J_{nn^{\prime}\bar{\sigma}}^{l}a_{nl\sigma}^{\dagger}a_{n^{\prime}l\sigma}\nonumber \\
 & + & \sum_{n,n^{\prime},l}F_{nn^{\prime}}^{l}a_{nl\uparrow}^{\dagger}a_{n^{\prime}l\downarrow}^{\dagger}+\mathrm{H.c}.\end{eqnarray}
Here the single particle energies are $\varepsilon_{nl}=\hbar\omega\left(2n+l+3/2\right)$,
and the Hartree interaction is described by the elements \[
J_{nn^{\prime}\sigma}^{l}=\int_{0}^{\infty}dr\ r^{2}R_{nl}(r)n_{\sigma}(r)R_{n^{\prime}l}(r),\]
and the pairing field is described by \[
F_{nn^{\prime}}^{l}=\int_{0}^{\infty}dr\ r^{2}R_{nl}(r)\Delta(r)R_{n^{\prime}l}(r).\]
We note that the $\sigma$ dependence of the Hartree term is due to
the population imbalance which implies that the corrections are different
for the two components. The density of $\sigma$ atoms is \begin{equation}
n_{\sigma}(r)=\sum_{n,n^{\prime},l}\frac{2l+1}{4\pi}R_{nl}(r)R_{n^{\prime}l}(r)\langle a_{nl\sigma}^{\dagger}a_{n^{\prime}l\sigma}\rangle,\end{equation}
 and the order parameter is \begin{equation}
\Delta(r)=\tilde{U}(r)\sum_{n,n^{\prime},l}\frac{2l+1}{4\pi}R_{nl}(r)R_{n^{\prime}l}(r)\langle a_{nl\uparrow}^{\dagger}a_{n^{\prime}l\downarrow}^{\dagger}\rangle.\end{equation}
The additional factor $(2l+1)$ comes from the degeneracy of the $l$-states,
obtained by the summation over the $m$ states. The renormalized interaction
strength $\tilde{U}(r)$ will be described later.

Truncating the Hilbert space by keeping only the states with single-particle
energies $\varepsilon_{nl}\leq E_{\mathrm{c}}$ allows writing the
Hamiltonian in a matrix form that can be diagonalized. Noticing that
the Hamiltonian commutes for different $(l)$-quantum numbers makes
it possible to diagonalize each $(l)$-matrix separately, simplifying
the problem significantly. For a given $(l)$-quantum numbers the
Hamiltonian reads \begin{equation}
H_{\mathrm{MF}}^{(l)}=\left(\begin{array}{c}
a_{0l\uparrow}^{\dagger}\\
\ldots\\
a_{N_{l}l\uparrow}^{\dagger}\\
a_{0l\downarrow}\\
\ldots\\
a_{N_{l}l\downarrow}\end{array}\right)^{T}M^{l}\left(\begin{array}{c}
a_{0l\uparrow}\\
\ldots\\
a_{N_{l}l\uparrow}\\
a_{0l\downarrow}^{\dagger}\\
\ldots\\
a_{N_{l}l\downarrow}^{\dagger}\end{array}\right),\end{equation}
where $N_{l}=[E/(\hbar\omega)-l-3/2]/2$ is the $l$-specific cutoff
(yielding the correct energy cutoff) and $M^{l}$ is a $2(N_{l}+1)\times2(N_{l}+1)$-dimensional
orthogonal matrix. Notice that the $\downarrow$ states have been
turned into holes by switching the order of $a_{\downarrow}^{\dagger}$
and $a_{\downarrow}$ operators as suggested by the ordinary Bogoliubov
transformation. The matrix $M^{l}$ is now \begin{widetext} \begin{equation}
M^{l}=\left(\begin{array}{cccccc}
\varepsilon_{0l}-\mu_{\uparrow}+UJ_{00\downarrow}^{l} & \ldots & UJ_{{0N}_{l}\downarrow}^{l} & F_{00}^{l} & \ldots & F_{{0N}_{l}}^{l}\\
\ldots & \ldots & \ldots & \ldots & \ldots & \ldots\\
UJ_{N_{l}0\downarrow}^{l} & \ldots & \varepsilon_{N_{l}l}-\mu_{\uparrow}+UJ_{N_{l}N_{l}\downarrow}^{l} & F_{N_{l}0}^{l} & \ldots & F_{N_{l}N_{l}}^{l}\\
F_{00}^{l} & \ldots & F_{N_{l}0}^{l} & -\varepsilon_{0l}+\mu_{\downarrow}-UJ_{00\uparrow}^{l} & \ldots & -UJ_{N_{l}0\uparrow}^{l}\\
\ldots & \ldots & \ldots & \ldots & \ldots & \ldots\\
F_{0N_{l}}^{l} & \ldots & F_{N_{l}N_{l}}^{l} & -UJ_{0N_{l}\uparrow}^{l} & \ldots & -\varepsilon_{N_{l}l}+\mu_{\downarrow}-UJ_{N_{l}N_{l}\uparrow}^{l}\end{array}\right).\end{equation}
 \end{widetext} Diagonalizing this Hamiltonian corresponds to the
Bogoliubov transformation, which yields the eigenenergies $E_{jl}$
and the corresponding quasiparticle eigenstates ($2(N_{l}+1)$-dimensional
vectors) $W_{jn}^{l}$.  The indices $j,n$ both run from $0$ to
${2N}_{l}+1.$ In the basis of these quasiparticle states, the Hamiltonian
is \begin{equation}
H_{MF}^{(l)}=\sum_{j=0}^{2N_{l}+1}E_{jl}\alpha_{jl}^{\dagger}\alpha_{jl}.\end{equation}
 In this basis, the density of atoms in $\uparrow$ state is  \begin{eqnarray}
n_{\uparrow}(r) & = & \sum_{l}\frac{2l+1}{4\pi}\sum_{j}\sum_{n,n^{\prime}=0}^{2N_{l}+1}R_{nl}(r)R_{n^{\prime}l}(r)\nonumber \\
 & \times & W_{jn}^{l}W_{jn'}^{l}n_{\mathrm{F}}(E_{jl}),\end{eqnarray}
 where the Fermi distribution $n_{\mathrm{F}}(E)=1/(1+e^{E/k_{\mathrm{B}}T})$.
Likewise, the density of atoms in $\downarrow$ state is \begin{equation}
\begin{split}n_{\downarrow}(r)=\sum_{l}\frac{2l+1}{4\pi} & \sum_{j}\sum_{n,n^{\prime}=0}^{2N_{l}+1}R_{nl}(r)R_{n^{\prime}l}(r)\\
 & \times W_{j,n+N_{l}+1}^{l}W_{j,n^{\prime}+N_{l}+1}^{l}n_{\mathrm{F}}(-E_{jl}),\end{split}
\end{equation}
 where the sign in the Fermi function is changed because of the $\downarrow$
component (the last $N_{l}+1$ components) of the   eigenstates correspond
to holes as compared to the particles in the  $\uparrow$ components
(the first $N_{l}+1$ components). The order parameter is given by\begin{eqnarray}
\Delta(r) & = & \sum_{l}\frac{2l+1}{4\pi}\sum_{j}\sum_{n,n^{\prime}=0}^{{2N}_{l}+1}R_{nl}(r)R_{n^{\prime}l}(r)\nonumber \\
 & \times & W_{jn}^{l}W_{j{,n}^{\prime}+N_{l}+1}^{l}\left(1+2n_{\mathrm{F}}(E_{jl})\right).\label{eq:finalgap}\end{eqnarray}
The total number of atoms in the two components can be obtained by
integrating over the densities. However, numerically it is faster
to calculate them directly from the eigenstates using the equations
\begin{equation}
N_{\uparrow}=\sum_{l}(2l+1)\sum_{j}\sum_{n=0}^{2N_{l}+1}W_{jn}^{l}W_{jn}^{l}n_{\mathrm{F}}(E_{jl})\label{eq:finalnumberup}\end{equation}
 and \begin{equation}
N_{\downarrow}=\sum_{l}(2l+1)\sum_{j}\sum_{n=0}^{2N_{l}+1}W_{j,n+N_{l}+1}^{l}W_{j,n+N_{l}+1}^{l}n_{\mathrm{F}}(-E_{jl}).\label{eq:finalnumberdown}\end{equation}
The gap and the number equations (\ref{eq:finalgap},\ref{eq:finalnumberup},\ref{eq:finalnumberdown})
are solved iteratively. We have made calculations where the Hartree
fields are included and compared them to the case where Hartree fields
are neglected by setting $J_{nn^{\prime}\sigma}^{l}=0$ for all $n,n^{\prime},l$
. Note that close to the Feshbach resonance, the Hartree fields become
formally infinite. In this extreme limit, we have limited the Hartree
interaction strength $U$ to $\left|\left(k_{\mathrm{F}}a_{s}\right)^{-1}\right|\leq\beta$,
where $\beta\approx0.5$. When comparing the results with and without
Hartree fields, there is no difference in the qualitative features
such as the order parameter oscillations and over-all shape of the
gap and density profiles, the Hartree fields cause only minor corrections
to gap and density profiles (effectively compressing the gas slightly).
Numerically, neglecting the Hartree fields gives a tremendous speedup
in the numerical solution because it decouples the density and gap
profiles. In the results presented in this article, the Hartree fields
are neglected. We have made the calculations at zero temperature and
present here only results at $T=0$. We have checked that the BdG
results do not change by using a finite but very small temperature
$T=0.001\ T_{\mathrm{F}}$.

As the density and gap profiles are decoupled, it is straightforward
to solve the gap equation for given chemical potentials. However,
since we want to keep the number of atoms fixed, the total procedure
will require optimizing also the chemical potentials, so that the
number equations are satisfied. The subsequent iteration procedure
can be performed in several different ways. We have found that a very
efficient procedure is to solve the chemical potentials (by solving
the number equations (\ref{eq:finalnumberup},\ref{eq:finalnumberdown}))
for each trial gap profile $\Delta(r)$. The trial gap profile $\Delta(r)$
is then used for solving the new trial gap profile $\Delta^{\prime}(r)$
using the gap equation (\ref{eq:finalgap}) with the new obtained
chemical potential. The chemical potentials therefore keep changing
between the iteration steps of the gap profile. On the other hand,
the numbers of atoms stay fixed in the iteration process. %
\begin{figure}
\begin{centering}\includegraphics[width=7cm]{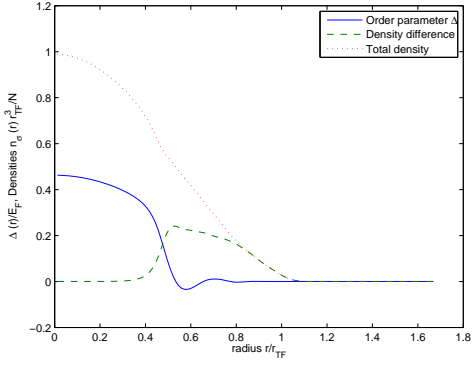} \par\end{centering}

\caption{(color online) Gap and density profiles on the BCS side of the resonance
$(k_{\mathrm{F}}a_{s})^{-1}=-0.5$ for $P=0.50.$ \label{fig:densityprofilesP500}}
\end{figure}
The initial guesses for the profiles needed for the iteration procedure
are obtained by using the chemical potentials and the gap profile
obtained for a slightly lower polarization and, eventually, by the
solution obtained for unpolarized gas. Solution of the profiles for
a given polarization $P$ requires therefore solving the profiles
for all lower polarizations. The validity of the final solution $\Delta(r),\mu_{\uparrow},\mu_{\downarrow}$
has been checked for several values of polarization and interaction
strength by perturbing the final solution and using the perturbed
profiles $\Delta'(r),\mu_{\uparrow},\mu_{\downarrow}$ as initial
guesses for the profiles. Usually the disturbed profiles have converged
either into the final profiles $\Delta(r)$ or into the normal state
$\Delta(r)\equiv0$. Only in the regime where our calculations predict
the superfluid core polarization have our solutions sometimes strayed
into a superfluid solution with unpolarized core. In such cases, the
final solution has been picked by choosing the solution with the lower
total energy, given by \begin{equation}
\begin{split}E=\sum_{l}^{2N_{l}+1}(2l+1) & \left[\sum_{n}\left(\varepsilon_{nl}-\mu_{\downarrow}\right)+\sum_{j}E_{jl}n_{\mathrm{F}}(E_{jl})\right]\\
 & -\frac{1}{\tilde{U}(r)}\int dr\ r^{2}|\Delta(r)|^{2},\end{split}
\label{eq:bdg-free-energy}\end{equation}
where the first (constant) term comes from switching to use the hole
states in the $\downarrow$ component (and this is countered by having
half of the eigenenergies $E_{jl}$ negative). The last term follows
from the mean-field term $\langle\hat{\Psi}_{\uparrow}^{\dagger}(\vec{r})\hat{\Psi}_{\downarrow}^{\dagger}(\vec{r})\rangle\langle\hat{\Psi}_{\downarrow}(\vec{r})\hat{\Psi}_{\uparrow}(\vec{r})\rangle$
which was dropped from the mean-field Hamiltonian because it is a
number, not operator.

The summation in the gap equation (\ref{eq:finalgap}) is divergent
without the energy cutoff $E_{\mathrm{c}}.$ This is a well known
phenomenon, following from the incapability of the contact interaction
potential to describe properly high energy behavior. Several different
regularization schemes have been proposed \cite{Bruun1999a,Bulgac2002,Grasso2003},
and here we apply the one suggested in Ref. \cite{Grasso2003}.  This
implies the following form for the renormalized coupling \[
\frac{1}{\tilde{U}(r)}=\frac{1}{U}+\frac{1}{2\pi^{2}}\left(\frac{k_{\mathrm{F}}^{0}(r)}{2}\ln\left(\frac{k_{c}(r)+k_{\mathrm{F}}^{0}(0)}{k_{c}(r)-k_{\mathrm{F}}^{0}(0)}\right)-k_{c}(r)\right),\]
where the momentum cutoff $k_{c}(r)=(2N_{c}+3-r^{2})^{1/2}$ and the
local Fermi momentum $k_{\mathrm{F}}^{0}(r)=(\mu_{\uparrow}+\mu_{\downarrow}-r^{2})^{1/2}$
for the imbalanced case compared to Eq. (14) and Eq. (18) of Ref.
\cite{Grasso2003}, respectively. On the BCS side of the resonance
for $\left(k_{\mathrm{F}}a_{s}\right)^{-1}<0$ we have used as the
cutoff $E_{\mathrm{c}}=200\ \hbar\omega$ and on the BEC side for
$\left(k_{\mathrm{F}}a_{s}\right)^{-1}>0$ the cutoff $E_{\mathrm{c}}=240\ \hbar\omega$
was used. We have tested the remaining cutoff dependence by using
the cutoff $300\ \hbar\omega.$ Depending on the number of atoms,
the cutoff dependence in the gap profiles was at most $2\%.$

\section{Results }

The results show two features that we will discuss in detail below:
1) For small and intermediate polarizations, FFLO-type oscillations
in the superfluid-normal interface at the edge of the trap, 2) for
large polarization, a polarized superfluid that extends through the
whole trapped gas. We study the behavior of these features, especially
1), throughout the BCS-BEC crossover and when the system size, i.e.
the atom number, is varied. We also calculate the condensate fraction
to make a connection to experiments. Finally, we analyze which harmonic
oscillator states are involved in pairing for different polarizations
and discuss the connection and differences to FFLO-state in a homogeneous
space. %
\begin{figure}
\begin{centering}\includegraphics[width=7cm]{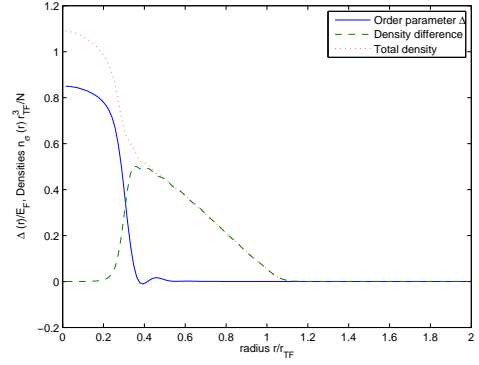} \par\end{centering}

\caption{(color online) Typical profiles at resonance $(k_{\mathrm{F}}a_{s})^{-1}=0.0$
for $P=0.9.$\label{fig:res-profiles} }
\end{figure}

\subsection{FFLO-type oscillations}

Typical density and gap profiles at $T=0$, for small polarization,
are shown in Figs. \ref{fig:comp-zero-T} and \ref{fig:dens-comp-zero-T},
both for LDA and BdG. Comparison of the profiles gives the following
general picture: BdG predicts a) SF (equal densities) core, b) PS
(FFLO-like oscillations) shell, c) normal state shell of the majority
component $N_{\uparrow}.$ LDA predicts a) SF (equal densities) core,
b) normal state shell; the absence of the PS shell in LDA is reflected
in the  discontinuity of the density and gap profiles at the SF-N
phase boundary. At finite temperatures, also LDA shows a polarized
shell and the boundary becomes continuous, only showing a kink (Fig.
\ref{fig:comp-finite-T}). As shown by Figs. \ref{fig:comp-zero-T}-\ref{fig:comp-finite-T},
for small polarization the PS given by BdG calculations is a narrow
shell and can be understood as a boundary effect. However, in the
  following we show that, for large polarization, the FFLO-features
extend to the center of the trap as well. 

A seen from Fig. \ref{fig:comp-zero-T} and Fig. \ref{fig:comp-finite-T}
the general results from LDA and BdG calculation agree fairly well
and the overall agreement for the gap and density profiles becomes
better for increasing particle number. The incapability of LDA to
correctly describe the short scale behavior (explicitly excluded by
LDA) leads to the unphysical appearance of a discontinous order parameter
solution and can be viewed as a breakdown of LDA near the interface
much in the same way as LDA generally breaks near the edge of a balanced
condensate. The breakdown of LDA is however more pronounced in the
imbalanced case due to presence of the normal shell surrounding the
condensate which leads to separation of the size of the condensate
and the size of the cloud. %
\begin{figure}
\begin{centering}\includegraphics[width=7cm]{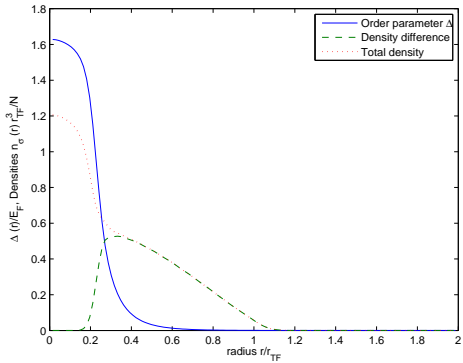} \par\end{centering}

\caption{(color online) Gap and density profiles on the BEC side of the resonance
$(k_{\mathrm{F}}a_{s})^{-1}=0.5$ for P=0.95. \label{fig:bec-profiles}}
\end{figure}

\subsubsection{Dependence on the interaction strength -- BCS-BEC crossover}

We present here results for three different cases: (1) BCS side of
the crossover, $(k_{\mathrm{F}}a_{s})^{-1}=-0.50$, (2) unitarity,
that is, $(k_{\mathrm{F}}a_{s})^{-1}=0$, (3) BEC side $(k_{\mathrm{F}}a_{s})^{-1}=0.50$.
We focus on the behavior at strong interactions since the cases (1)-(3)
can all be considered to be at the unitarity regime (if it is defined
$|(k_{\mathrm{F}}a_{s})^{-1}|<1$ ) although representing different
sides of the crossover. We do not consider the extreme BCS limit of
weak interactions because the features and trends observed in the
unitarity limit are also expected to appear at weaker coupling. 

For attractive interactions, $(k_{\mathrm{F}}a)^{-1}<0$, the typical
density and order parameter profiles looks as shown in Fig. \ref{fig:densityprofilesP500}
for polarization $P=0.50$. In agreement with earlier studies using
the same approach \cite{Castorina2005,Mizushima2005a,Kinnunen2006a,Machida2006},
the solution reveals an unpolarized BCS-type region at the center
of the trap and a polarized shell with oscillating order parameter.
The oscillations rapidly dampen when the density of the minority component
drops.

We have studied the presence of the oscillations around the unitarity
region, and noticed that the critical polarization for the appearance
of the order parameter oscillations increases with stronger interactions.
At unitarity ($(k_{\mathrm{F}}a)^{-1}=0.0$), the order parameter
oscillations did not appear until polarization $P=0.70$, compared
to $P=0.10$ at $(k_{\mathrm{F}}a)^{-1}=-0.50$. The disappearance
of the oscillations follows from the enhanced stability of the BCS-type
pairing at stronger interactions, making the increased distortion
of the minority component density profile favorable over the reduced
order parameter due to the polarization. The result at unitarity is
shown in Fig. \ref{fig:res-profiles}. 

On the BEC side, the oscillations do not appear. We have made calculations
for several parameters on the BEC side of the resonance and have not
found any FFLO-type oscillations in the order parameter on the BEC
side. This is consistent with the observation discussed above that
the critical polarization for the emergence of the nodes in the order
parameter increases with increasing interaction strength. The same
qualitative results on the BCS-BEC crossover were recently obtained
in \cite{Mizushima2007} using a hybrid BdG-LDA scheme in a cylindrically
symmetric geometry. Typical density and gap profiles on the BEC side
are shown in Fig. \ref{fig:bec-profiles}. %
\begin{figure}
\begin{centering}\includegraphics[width=7cm]{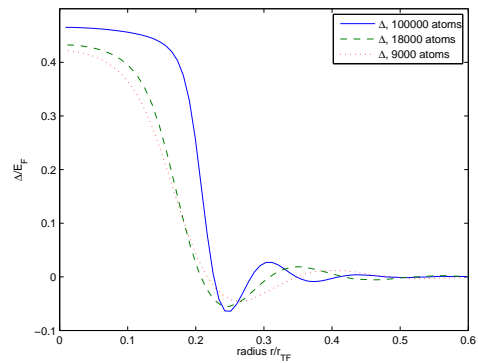} \par\end{centering}

\caption{(color online) Gap profiles at $(k_{\mathrm{F}}a_{s})^{-1}=-0.5$
for several values $N=9000,18000,100000$ of the total number of particles
at a high and fixed polarization $P=0.74.$ For increasing particle
number the transition from an almost constant value of the gap in
the BCS core to the oscillating gap at the edge becomes sharper. For
increasing particle number the gap FFLO-like oscillations at the edge
the cloud display shorter wavelength and a slight increase in amplitude.
\label{fig:gapsP740}}
\end{figure}

\subsubsection{Dependence on the system size - atom number N}

\begin{figure}
\begin{centering}\includegraphics[width=7cm]{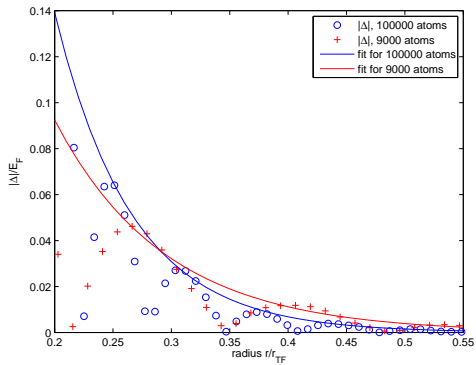} \par\end{centering}

\caption{(color online) Exponential fits for the envelopes of the order parameter
oscillations for 9000 and 100000 atoms. \label{fig:expfitfortail}}
\end{figure}
We have studied the dependence of the results on the number of atoms
by performing the BdG calculations for the total atom numbers 9000,
10000, 18000, 20000 and 100000. Note that this means substantial increase
in the system size compared to our earlier work \cite{Kinnunen2006a}
where the atom number 10000 was used. 

The order parameter oscillations for different numbers of atoms are
shown in Fig. \ref{fig:gapsP740}. In Fermi units (i.e. in the cloud
scale $R_{\mathrm{TF}}$), the wavelength of the oscillations becomes
shorter with the increased atom number but, on the other hand, more
nodes appear. The scaling of all these factors is complicated, but
insight can be obtained by fitting an exponential function $\sim e^{-x/\xi}$
to the envelope of the order parameter oscillations as shown in Fig.
\ref{fig:expfitfortail}. The exponential decay gives a very good
description for the damping of the oscillations with increasing distance
from the trap center $r$. The penetration depths $\xi$ obtained
from the fitting indeed slowly decrease in Fermi units when the atom
number $N$ is increased, the scaling being roughly $N^{1/6}$. Since
$R_{\mathrm{TF}}\propto N^{1/6}$, one may anticipate that the decrease
of the penetration depth only occurs in relative scale, not absolute.
Indeed, in a microscopic length scale (harmonic oscillator units $a_{\mathrm{osc}}$)
the penetration depth remains constant, being $0.54\ a_{\mathrm{osc}}$
for 100000 atoms and $0.56\ a_{\mathrm{osc}}$ for 9000 atoms. These
observations confirm that the order parameter oscillations are not
a finite size effect but rather an interface effect for the superfluid-normal
interface. From the shape of the order parameter profile it can be
seen that that the interface formed becomes sharper for increasing
particle number and we therefore suggest that the interface forming
at the edge is an analog of the proximity effects appearing in superfluid-normal
and superfluid-ferromagnetic junctions. In the latter case it has
been shown that the interplay between BCS superconductivity and ferromagnetic
order (in our case the polarization $\delta n$) gives rise to an
oscillating order parameter which decays anomalously slow (over several
oscillations) on the ferromagnet side and with a characteristic length
scale that is independent of the properties of the superconductor
\cite{Demler1997,Kontos2001,Schaeybroeck2007}. An interesting question
is the dependence of the penetration depth on the interaction strength
and polarization.  However, we have not been able to pursue this
question in more detail here and leave it for future work.

\begin{figure}
\includegraphics[width=7cm]{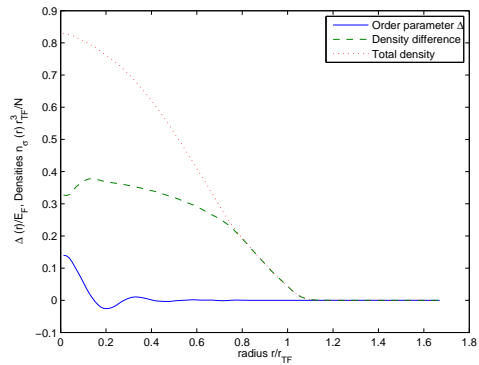}

\caption{Density and order parameter profiles at $P=0.762$ and $(k_{\mathrm{F}}a_{s})^{-1}=-0.50$
for $N=9000$ show the superfluid with core polarization, together with
a small dip in the density difference at the center of the trap. }

\label{fig:densityprofilesP762} 
\end{figure}
In this context, we believe a few remarks on the work \cite{Liu2007}
are in order. It is argued in \cite{Liu2007}, based on a hybrid BdG-LDA
scheme (no analysis of the penetration depths as presented above was
done in \cite{Liu2007}), that the oscillations of the order parameter
vanish for sufficiently large particle number and are thus interpreted
as a finite size effect. It is argued that as the cutoff energy $E_{c}$
introduced to separate the BdG and the LDA scales in the hybrid scheme
goes to zero, LDA should be recovered. This is of course self-evident
due to the construction of the hybrid algorithm but reducing the cutoff
energy without concern may lead to missing some features of the system.
The fact that a sharp interface with increasingly microscopic features
(oscillations with short period on the scale of the cloud size) builds
up for increasing number of particles requires $E_{c}$ to be increased
significantly in order to resolve the short length scale features.
 Consistent with our results, Fig. 4 in \cite{Liu2007} shows that
the order parameter oscillates with a shorter period, and at the same
time the amplitude of the order parameter slowly increases, for increasing
particle number. The recent results for a hybrid BdG-LDA scheme \cite{Mizushima2007}
and the scaling analysis therein agree well with the results given
by our full BdG calculations.

\subsection{Polarized superfluid for large imbalances}

When the polarization $P$ exceeds some (large) critical value, the
FFLO-type oscillations discussed above reach the center of the trap.
At this point the gas becomes polarized also at the center of the
trap and the order parameter drops to roughly half of its unpolarized
BCS value. This realizes a superfluid with finite polarization throughout
the system. Fig. \ref{fig:densityprofilesP762} shows the density
and order parameter profiles for $P=0.762$. For such a core polarized
system, the concept of an interface effect is intriguing as there
is no BCS-type superfluid core present. 

The density difference for such a polarized superfluid shows an interesting
feature: a small dip in the center of the trap (i.e. the density difference
is smaller than in the surrounding area, but still non-zero). At zero
temperature, this feature only appears in our BdG calculations whereas
LDA, which fails to predict a polarized superfluid at $T=0$, does
not lead to such a dip in the density difference. In contrast, LDA
calculations at finite temperature produce such a feature in connection
with the finite temperature BP phase. Recent Monte Carlo studies of
the trapped Fermi gas have shown that, for the strongly interacting
normal state, the density difference increases monotonously towards
the center of the trap \cite{Lobo2006}. Therefore one may argue that
the dip is associated with a polarized superfluid: at $T=0$ it is
FFLO-type, at temperatures that are finite but clearly below the critical
temperature it is either FFLO or BP. At higher temperatures, pseudogap
effects may contribute to such features as well \cite{Chien2006b}.
Such a dip can be seen in the experimental results of \cite{Shin2006a}.

The interaction strength dependence of such a polarized superfluid
with oscillating order parameter is similar to what was already discussed
above. Since there are no order parameter oscillations on the BEC
side, we have also not seen any core polarized superfluid either.
Of course, there does exist a different kind of core polarization
on the BEC side: coexistence of a molecular condensate and free excess
fermions, but that is not associated with oscillating order parameter.

The parameter window for the polarized superfluid shrinks with increasing
atom number $N$. For $9000$ atoms the window is $0.746<P<0.784,$
whereas for $18000$ atoms it is $0.746<P<0.774.$ Since the convergence
of the calculations near this critical window is slow, we have not
been able to determine the corresponding window for $100000$ and
have not systematically analyzed how the window scales with particle
number. For the order parameter oscillations, we have shown in section
III.A.2. that their absolute length scale stays unchanged for large
particle numbers. Based on the present data, we cannot conclude with
similar confidence whether the parameter window for the existence
of the core polarized superfluid becomes negligible or not for large
condensates. However, we would like to emphasize that the core polarized
superfluid is not due to a trivial finite size effect, i.e. not originating
from having discrete oscillator states in the system description.
As seen in Refs. \cite{Grasso2003,Machida2006}, such finite size
effects manifest as a narrow dip in the density and order parameter
profiles at the center of the trap. However, these effects vanish
when the number of atoms increases or when the interaction strength
is increased. Because of the stronger interactions, we do not see
these finite size effects even at $9000$ atoms. Note that such a
dip originating from finite size effects is completely different from
the dip in the density difference discussed above as a signature of
the core polarized superfluid.

\subsection{Condensate fraction}

\begin{figure}
\includegraphics[width=7cm]{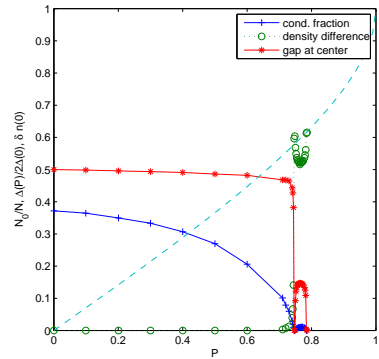}

\caption{The condensate fraction, density difference, and the order parameter
as a function of polarization $P$ obtained for $9000$ atoms. The
interaction strength is $(k_{\mathrm{F}}a)^{-1}=-0.50$.}

\label{fig:condfracN9000} 
\end{figure}
\begin{figure}
\includegraphics[width=7cm]{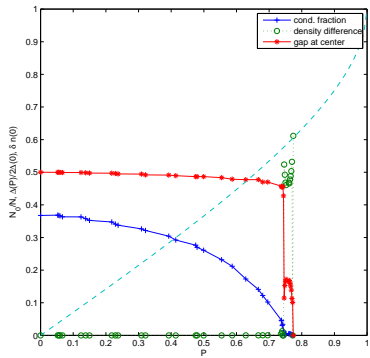}

\caption{The condensate fraction, density difference, and the order parameter
as a function of polarization $P$ obtained for $18000$ atoms. The
interaction strength is $(k_{\mathrm{F}}a)^{-1}=-0.50$.}

\label{fig:condfracN18000} 
\end{figure}

To make connection to experiments \cite{Shin2006a} where condensate
fractions are measured, we calculate here the condensate fraction
for an imbalanced gas. In the case of a balanced Fermi gas the condensate
fraction is defined in \cite{Campbell1997,Salasnich2005} and can
be viewed as a measure of the number of condensed pairs which in the
extreme BEC limit and at zero temperature is just $N/2\equiv N_{\sigma}.$
For the imbalanced gas,  the corresponding number of molecules in
the asymptotic BEC limit is the number of minority atoms $N_{\downarrow}.$
It is therefore natural to consider the following normalized condensate
fraction \begin{equation}
N_{0}/N_{\downarrow}=\frac{1}{N_{\downarrow}}\int d^{3}\vec{r}_{1}d^{3}\vec{r}_{2}\left|\langle\Psi_{\uparrow}(\vec{r}_{1})\Psi_{\downarrow}(\vec{r}_{2})\rangle\right|^{2},\label{eq:imbalance-cf}\end{equation}
for the condensate fraction in the case of $N_{\downarrow}<N_{\uparrow}.$
Figs. \ref{fig:condfracN9000} and \ref{fig:condfracN18000} show
the density difference, the order parameter at the center of the trap,
and the imbalanced condensate fraction from Eq. \prettyref{eq:imbalance-cf}
as function of the polarization for $(k_{\mathrm{F}}a_{s})^{-1}=-0.50$.
The critical polarization is roughly $P_{c}=0.78$ and the transition
is reflected in a rapid increase  in the density difference in the
center of the trap. Also the order parameter in the center of the
trap drops rapidly. However, as the figures show, there exists a narrow
but finite polarization region (in this case for $0.75<P<0.78$) where
both the density difference and the gap have non-negligible values
in the center of the trap. All data points are for fully converged
order parameter and density profiles, satisfying the gap equation
with precision of $10^{-6}$. In addition, several points in the plot
have been calculated with different initial conditions for the iteration. 

\begin{figure}
\includegraphics[width=7cm]{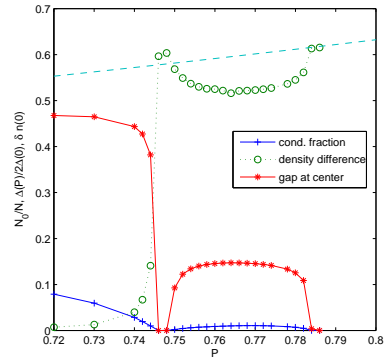}

\caption{The core polarized superfluid region for $9000$ atoms $(k_{\mathrm{F}}a_{s})^{-1}=-0.50.$}

\label{fig:condfracN9000detail} 
\end{figure}

The figures are in good qualitative agreement with the experimental
condensate fractions \cite{Shin2006a}, showing the sudden onset of
the density difference at the center of the trap when the condensate
fraction drops to zero. The core polarized superfluid manifests itself
as a weak revival of the condensate fraction as shown in Fig. \ref{fig:condfracN9000detail}.
In other words, a finite condensate fraction co-exists with a finite
central density difference. Although the qualitative agreement with
the experimental results in \cite{Shin2006a} is good, higher experimental
accuracy as well as extending our calculations to finite temperatures
is needed for quantitative comparison, especially regarding the small
window for the core polarized superfluid. Since the condensate fraction
is small in the interesting transition region, better signatures of
the polarized FFLO-type superfluid may be provided by the central
gap, and by the dip in the central density difference as discussed
above, provided that temperature dependence is also carefully investigated.

\subsection{Contribution of different harmonic oscillator states in pairing}

\begin{figure}
\begin{centering}\includegraphics[width=7cm]{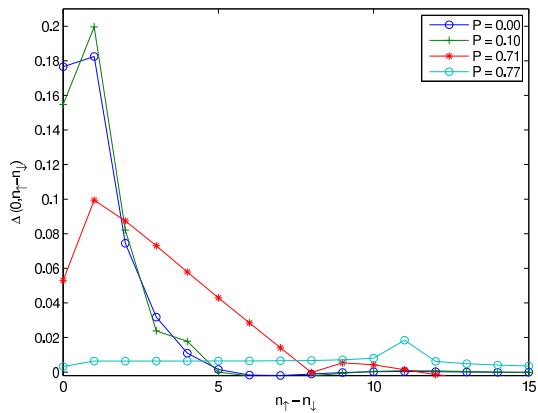}\par\end{centering}

\caption{(color online) The gap at the center of the trap (the part of the
gap for the quantum number $l=0$ ) as function of the difference
in radial quantum number $\Delta n$ for different values of the polarization
and for $N=18000.$ For increasing polarization the mainly intra-shell
and nearest-neighbor-shell pairing indicated by the central peak (at
$P=0$ ) diminishes and a secondary peak appears for large $\Delta n$
which indicate the increasing importance of inter-shell pairing between
shell states having an energy separation of order determined by the
mismatch of Fermi energies. \label{fig:pairingatcenter} }
\end{figure}

We studied the origin of the oscillations  in the BdG order parameter
by considering which trap shells (quantum number $n$) are involved
in pairing in the center of the trap, i.e.\ for quantum number $l=0$.
As shown in Fig. \ref{fig:pairingatcenter}, for zero polarization
$P=0$, the pairing involves 0-2 neighboring shells, i.e.\ is peaked
at $n-n'=0$, with considerable weight until $n-n'=2$ due to strong
interactions. For small polarization $P=0.1$ the peak is shifted,
but the weight is still very much on the same $n-n'$ as for $P=0$,
which results to only minor modification of the order parameter profile.
However, for large polarization $P\simeq0.75$, when the oscillations
of the order parameter appear also in the center of the trap, the
pairing is peaked at nonzero $n-n'\sim11$. Therefore, for large polarizations
the state clearly resembles the FFLO state where pairing is predominantly
between momentum states $k-k'$ determined by the mismatch of the
Fermi surfaces (in our example $P\simeq0.75$ means a mismatch of
the Fermi surfaces of about $n-n'\sim11$). We have found that this
behavior is not due to small particle number, in contrast, it becomes
more clear for larger particle numbers.

\section{Conclusions and discussion}

In summary, we have shown that FFLO-type oscillations appear in a
trapped polarized Fermi-gas and, for large polarization, extend to
the center of the trap thereby realizing a non-BCS superfluid at zero
temperature. These results were obtained by BdG calculations using
harmonic oscillator eigenstates and particle numbers up to 100000.
We have also made a comparison to the results given by local density
approximation.

The FFLO-type oscillations of the order parameter appear on the BCS
side of the BCS-BEC crossover. When the interaction strength is increased,
the polarization required for the existence of the oscillations grows.
On the BEC side, no oscillating order parameter was found. The scaling
analysis for the penetration depth of the oscillations shows that
their characteristic length scale stays constant when the particle
number is increased. Therefore the characteristic length scale relative
to the atom cloud scale (given by $R_{\mathrm{TF}}$) decreases very
slowly, $\propto N^{1/6}$. The features presented here should thus
be observable even for condensate sizes corresponding to present-day
experiments. RF-spectroscopy was proposed in our earlier work \cite{Kinnunen2006a}
for detecting the gapless excitations related to the nodes of the
order parameter as a signature of the FFLO-type oscillations. The
first experiments using RF-spectroscopy in investigating imbalanced
gases were recently done \cite{Schunck2007}. Moreover, the dip in
the central density difference could provide a signature of the core
polarized superfluid, as well as the simultaneous measurement of the
gap and the density difference in the center of the trap. 

The results presented here have an interesting connection to superconductor-ferromagnet
interface effects. For future work, it is fascinating to think about
the freedom that the ultracold gases offer in terms of designable
trapping geometries and other parameters: one should be able to systematically
study this kind of effects from the limit of having large superfluid
and polarized normal state (``ferromagnet'') regions and a small interface,
to the limit where the interface, superfluid, and normal state are
all of comparable size and novel mesoscopic phenomena may be found.

\begin{acknowledgments}
We thank W. Ketterle and M. Zwierlein for useful discussions. This
work was supported by Academy of Finland (project numbers 213362,
106299, 205470) and conducted as part of a EURYI scheme award. See
www.esf.org/euryi. J.K. acknowledges the support of the Department
of Energy, Office of Basic Energy Sciences via the Chemical Sciences,
Geosciences, and Biosciences Division.
\end{acknowledgments}
\bibliographystyle{apsrev} \bibliographystyle{apsrev} \bibliographystyle{apsrev}
\bibliography{pra}

\end{document}